\documentclass{elsart}

\usepackage{psfig}

\usepackage{natbib}

\begin{document}

\runauthor{Cicero, Caesar and Vergil}

% -------------------------------------------------------------------------

\begin{frontmatter}

\title{The contribution of Narrow-Line Seyfert 1 galaxies to the soft X-ray background}

\author[AIP]{G. Hasinger}
\author[AIP,PSU]{I. Lehmann}
\author[Cal]{M. Schmidt}
\author[Pri]{J.E. Gunn}
\author[PSU]{D.P. Schneider}
\author[AUI]{R. Giacconi}
\author[MPE]{J. Tr\"umper}
%\author[Bol,BolII]{G. Zamorani}
\author[Bol]{G. Zamorani}
\address[AIP]{Astrophysikalisches Institut Potsdam, An der Sternwarte 16,
D-14482 Potsdam, Germany}
\address[PSU]{Department of Astronomy \& Astrophysics, 525 Davey Lab, 
The Pennsylvania State University, University Park, Pennsylvania, PA 16802, USA}
\address[Cal]{California Institute of Technology, Pasadena, CA 91125, USA}
\address[Pri]{Princeton University Observatory, Peyton Hall, Princeton, NJ 08540, USA}
\address[AUI]{Associated Universities, Inc. 1400 16th Street, NW, Suite 730 Washington, DC 20036, USA}
\address[MPE]{Max-Planck-Institute f\"ur extraterrestrische Physik, Karl-Schwarzschild-Str. 1, 85748 
Garching bei M\"unchen, Germany}
\address[Bol]{Osservatorio Astronomico, Via Ranzani 1, I-40127, Bologna, Italy}
%\address[BolII]{Istituto di Radioastronomia del CNR, via Gobetti 101, I-40129, Bologna, Italy}

\begin{abstract}
The ROSAT Ultradeep HRI survey in the Lockman Hole contains a complete sample 
of 91 X-ray sources with 
fluxes in the 0.5-2 keV band larger than 1.2 $\cdot$ 10$^{-15}$ erg cm$^{-2}$
s$^{-1}$, where over ~75 \% of the sources are quasars or Seyfert galaxies.
During the course of our optical identification work, we have obtained optical
spectra of 67 narrow emission line galaxies (NELG), which are physically not
associated with the X--ray sources. We have derived the equivalent width (EW)
and the full width at half maximum (FWHM) for the most prominent emission
lines of 41 quasars and Seyfert galaxies taken from the ROSAT Deep Survey
(RDS), which has a flux limit of 5.5~$\cdot$~10$^{-15}$ erg cm$^{-2}$ s$^{-1}$
in the 0.5-2.0 keV band. Furthermore we have obtained the EW and FWHM values
of the field NELGs.  Here we present the spectroscopic discrimination between
RDS Seyfert galaxies and field galaxies (NELG). The analysis of the emission
lines has revealed that a single object out of 69 spectros\-copically
identified AGN fits the optical criteria of Narrow-Line Seyfert 1 galaxies
(NLS1).   This may indicate that NLS1 contribute only marginally to the soft
X-ray background, but we can not exclude a possible larger contribution.
\end{abstract}


\end{frontmatter}

% -------------------------------------------------------------------------

\section{The ROSAT Deep Surveys in the Lockman Hole}

The most sensitive ROSAT surveys consist of a 207 ksec ROSAT
PSPC exposure, a 205 ksec HRI raster scan and a total 1112 ksec HRI expsoure
of a $0.3\ deg^2$ area in the Lockman Hole region. The HRI images are the
basis  for the Ultradeep HRI Survey (Hasinger at al. 1999). At a flux limit of
10$^{-15}$~erg~s$^{-1}$ in the 0.5-2.0 keV energy band, the HRI survey has
resolved about 70-80\% of the soft X-ray background into discrete sources. The
ROSAT Deep Survey (RDS), based on the PSPC image, includes a statistically
complete sample of 50 X-ray sources with fluxes in the 0.5-2.0 keV band greater
than 5.5$\cdot$10$^{-15}$~erg~s$^{-1}$ (Hasinger et al. 1998).  The
spectroscopic identification of the RDS using the Keck and Palomar telescopes
have shown that about 75\% of the sources are quasars and Seyfert galaxies
(Schmidt et al. 1998, Lehmann et al. 2000). 

Both surveys contain 91 X-ray sources, where the faintest sources reach a flux 
of 1.2$\cdot$10$^{-15}$ erg s$^{-1}$ in the 0.5-2.0 keV band.
Recent optical/infrared work has led to an identification of 88 of the 91 
X-ray sources, confirming a high fraction of AGNs (72 objects). This is the
largest fraction of AGNs found in any previous X-ray survey (Boyle et al.
1995, Georgantopolous et al. 1996, Bower et a. 1996 and McHardy et al. 1998).
Among our AGNs is the most distant X-ray selected quasar at a redshift of 4.45
(Schneider et al. 1998).   Groups and clusters of galaxies ($\sim$10\%) form
the second most abundant class of objects. One X-ray source has been
classified as a narrow emission line galaxy (NELG), whereas some deep ROSAT
PSPC surveys found a significantly larger fraction of NELGs (Boyle et al.
1995, McHardy et al. 1998). We see no evidence that NELGs or other classes of
objects dominate the soft X-ray counts at faint fluxes. 

\section{Optical spectral properties of RDS quasars and Seyfert galaxies}

In the following we discuss the emission line properties of the 41 spectroscopically 
confirmed RDS quasars and Seyfert galaxies. 33 X-ray sources have been
identified with quasars and Seyfert 1 galaxies in the redshift range between
0.08 and 2.83. Their optical spectra show broad emission lines (FWHM $>$ 1500
km s$^{-1}$) of Ly$\alpha$ $\lambda$1216, C IV $\lambda$1548, C III]
$\lambda$1908 and Mg II $\lambda$2798 at medium and large redshifts, or of
Balmer lines (H$\alpha$ $\lambda$6563, H$\beta$4861) at lower redshifts. Most
of them can be classified as type I AGN, whereas the large Balmer decrement
of two Seyfert galaxies having broad H$\alpha$ or H$\beta$ emission lines
indicates type II AGN (see object 59A in Fig. 1). 

\begin{figure}[t]
\begin{center}
\begin{minipage}{13.8cm}
\begin{minipage}{6.7cm}
\psfig{figure=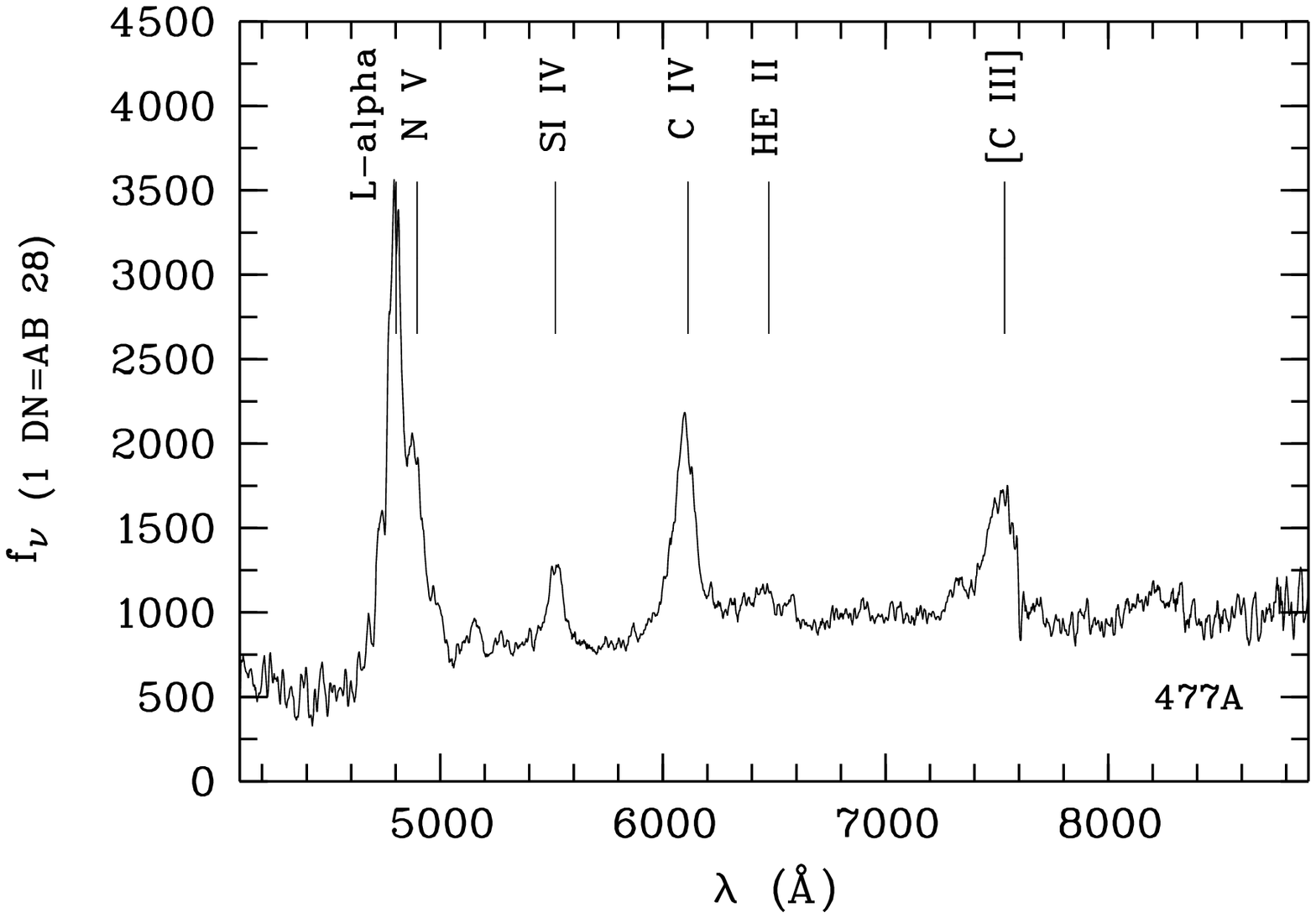,bbllx=45pt,bblly=52pt,bburx=403pt,bbury=566pt,width=6.7cm,angle=-90}
\end{minipage}
\hfill
\begin{minipage}{6.7cm}
\psfig{figure=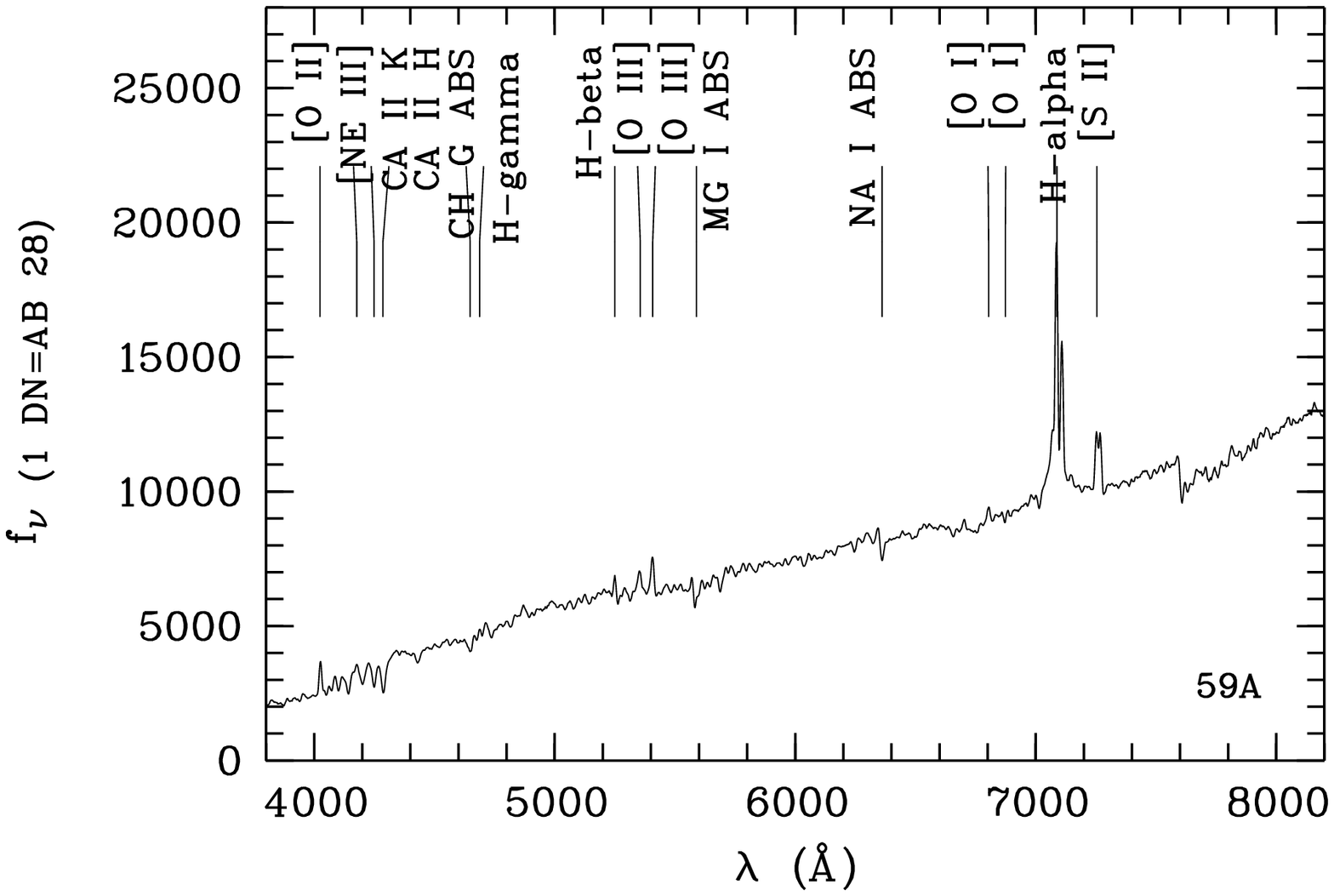,bbllx=45pt,bblly=52pt,bburx=403pt,bbury=566pt,width=6.7cm,angle=-90}
\end{minipage}
\end{minipage}
\end{center}
\caption{Keck LRIS spectra of the quasar 477A ($z=2.949$) and of the Seyfert 1.5 
galaxy 59A ($z=0.080$), which shows only a broad H${\alpha}$ emission line
(FWHM $\sim$ 8300 km s$^{-1}$.)} \end{figure}

The optical spectra of six RDS AGNs show only narrow emission lines with 
FWHM $<$ 1500 km s$^{-1}$ indicating a type II AGN. When their redshifts are
above 0.5, the spectra do not allow a classification using the diagnostic
diagrams of Osterbrock (1981). But the existence of high excitation [Ne~V]
$\lambda$3426 (see 26A in Fig. 2) or strong [Ne III]~$\lambda$3868 narrow
emission lines, together with an X-ray luminosity above 10$^{43}$ erg s$^{-1}$
in the 0.5-2.0 keV band, reveal an AGN (Schmidt et al. 1998).  Two further
sources have been classified as type II AGNs, because of their high X-ray
luminosity (log L$_{X}>43$). One of them is a radio galaxy at $z=0.708$
showing only typical galaxy absorption lines. Several RDS AGNs (type I and II)
below $z=1$ show a significant continuum emission originating in the host
galaxy (Fig. 1/Fig. 2). The RDS sample contains in total 35 type I and 6 type
II AGNs. The large $R-K^{\prime}$ colour of two spectroscopically unidentified
sources indicates either obscured AGNs or high-redshift clusters of galaxies
(Lehmann et al. 2000).

\section{Spectroscopic discrimination of Seyfert galaxies and field NELGs} 

In the course of the spectrocopic identification of our X-ray sources we have 
taken optical spectra of 83 field galaxies. 67 of them are narrow emission
line galaxies (NELGs) with no physical connection to the X-ray sources. There
is no soft X--ray emission (0.5-2.0 keV band) detected above a limiting flux
of 10$^{-15}$ erg cm$^{-2}$ s$^{-1}$ associated with these sources. The
optical spectra of the field NELG (cf. 14C in Fig. 2) show only narrow
emission lines, but no strong [Ne~III]~$\lambda$3868 or high ionization [Ne~V]
$\lambda$3426 emission lines. The possible misidentification of faint X-ray
sources with NELGs is discussed by Schmidt et al. (1998) and Lehmann et al.
(2000).

\begin{figure}[t]
\begin{center}
\begin{minipage}{13.8cm}
\begin{minipage}{6.7cm}
\psfig{figure=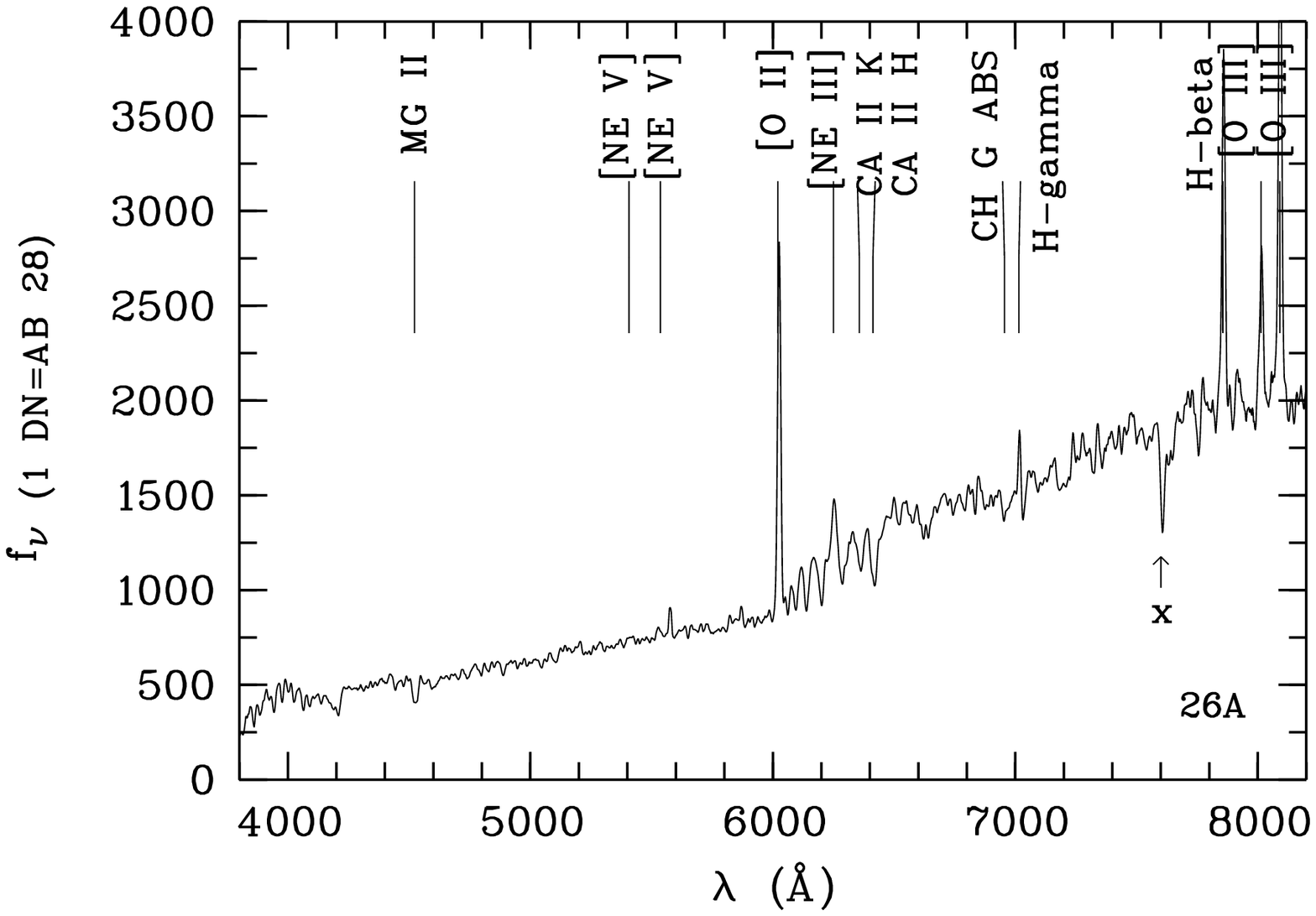,bbllx=45pt,bblly=52pt,bburx=403pt,bbury=566pt,width=6.7cm,angle=-90}
\end{minipage}
\hfill
\begin{minipage}{6.7cm}
\psfig{figure=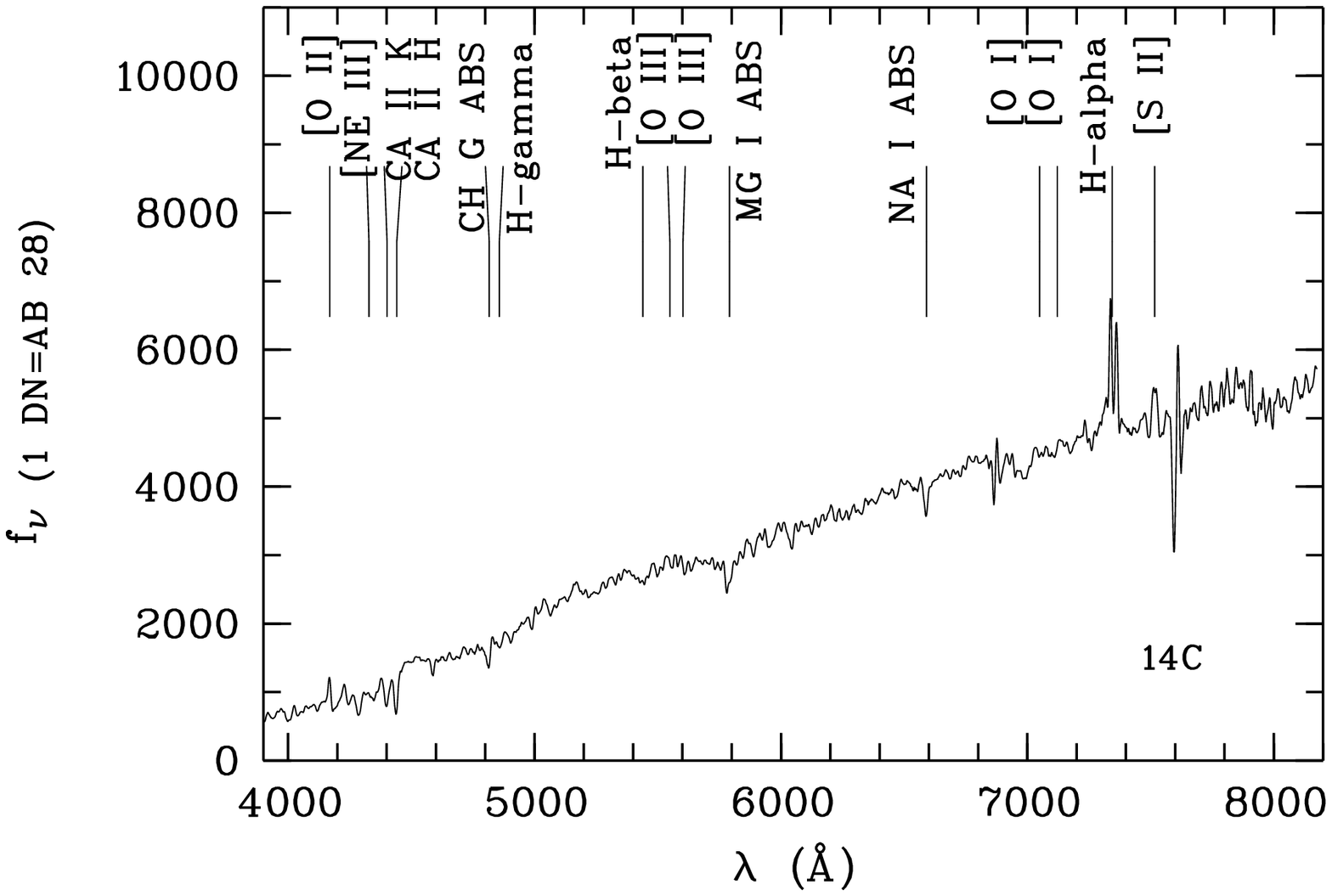,bbllx=45pt,bblly=52pt,bburx=403pt,bbury=566pt,width=6.7cm,angle=-90}
\end{minipage}
\end{minipage}
\end{center}
\caption{Keck LRIS spectra of the narrow emission line AGN 26A ($z=0.616$) 
from the RDS sample and the narrow emission line field galaxy (NELG) 14C at
$z=0.119$. The spectra of 26A and the NELG show a prominent galaxy continuum
above the Ca H$+$K$~\lambda\lambda$3934/3968 break. The high-ionization
[Ne~V]$ \lambda$3426 emission line in the spectrum of 26A confirms the AGN
nature of this object.} \end{figure}

To study the emission line properties we have derived the FWHM and the rest 
frame EW for the most prominent emission lines of the RDS quasars/Seyfert
galaxies and of the field NELGs. The mean FWHM/EW values for the broad
emission lines of RDS quasars/Seyfert 1 galaxies are consistent with those
values found for other X-ray selected AGN samples at lower mean redshift, eg.,
the RIXOS-sample (Puchnarewicz et al. 1997) and the CRSS-sample (Boyle et al.
1997), and the mean values derived from several emisssion line and UV/optical
selected AGN samples (Schmidt et al. 1986).

\begin{figure}[h]
\begin{minipage}{13.8cm}
\psfig{figure=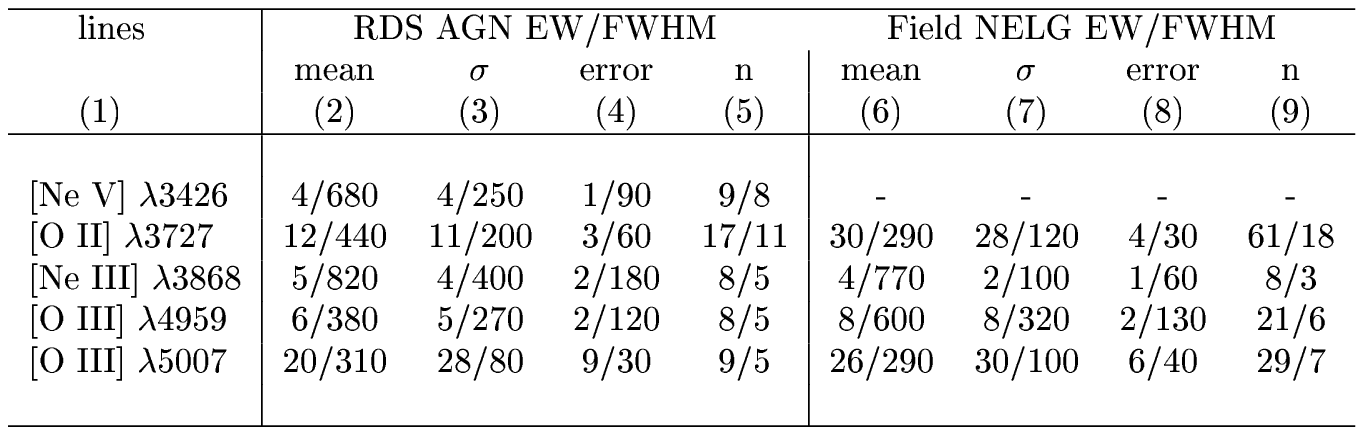,bbllx=70pt,bblly=350pt,bburx=462pt,bbury=496pt,width=13.8cm}
\end{minipage}
{\bf Tab.~1.} Rest frame EW [\AA ] and FWHM [km s$^{-1}$] values of the forbidden narrow 
emission lines of the RDS AGN in comparison with those of the field NELGs.
High ionization [Ne V] $\lambda$3426 emission lines have been detected only in
AGN. \end{figure}

The mean RDS EW values of the forbidden narrow emission lines, found for Seyfert galaxies 
at redshifts below 1, are slightly smaller than those from the AGN comparison
samples (Lehmann et al. 2000). The RDS EW of [Ne~III] $\lambda$3868 and the
[O~III] $\lambda \lambda$4959/5007 lines seem to be in better agreement with
those from field NELGs (Tab.~1) than with those from other AGN samples. The
optical spectra of most RDS AGNs, which contain forbidden narrow emission
lines, show several typical galaxy absorption lines, eg., Ca H$+$K $\lambda
\lambda$3934/3968, CH~G $\lambda$4301 (see 26A in Fig. 2). A strong continuum
contribution produced by the host galaxy could account for these less stronger
forbidden lines in our RDS AGNs.

\section{Contribution of NLS1 to the soft X-ray background}

Although the majority of our faint X-ray sources have been spectroscopically
identified  with quasars and Seyfert galaxies, only one out of 69 AGNs matches
the optical criteria of narrow line Seyfert 1 galaxies taken from Osterbrock \&
Pogge (1985) and more precisely defined by Goodrich (1989). These somewhat
subjective criteria are as follows. 1) NLS1 have slightly broader Balmer lines
($<$ 2000 km s$^{-1}$) in comparison to the forbidden lines such as [O~II]
$\lambda$3727 or [O~III] $\lambda \lambda$4959/5007. 2) The flux ratio [O~III]
$ \lambda$5007 to H$\beta$ $\lambda$4861 is $<$ 3, a level which allows to
discriminate between NLS1 and Seyfert 2 galaxies (Shuder \& Osterbrock 1981).
3) Emission lines of Fe II or of high ionization [Fe VII] $\lambda$6087 and
[Fe X] $\lambda$6375 are often present.

The spectrum of the NLS1 37A (Fig. 3) shows prominent Fe II bumps at 4500--4600 \AA~and 
at 5250--5350 \AA~(in the rest frame). The FWHM of the broad component of the
H$\beta$ $\lambda$4861 emission line is slightly above the limit for NLS1
(2070$\pm$20 km s$^{-1}$). The flux ratio [O~III] $\lambda$5007/H$\beta$
$\lambda$4861 is clearly below 3. The NLS1 37A at $z=0.462$ belongs to the
highest-redshift objects of this class. The bright soft X--ray selected sample
of Grupe et al. (1999) contains for example only 4 objects (12\%) above
$z=0.3$.

\begin{figure}[htb]
\begin{center}
\begin{minipage}{6.7cm}
\psfig{figure=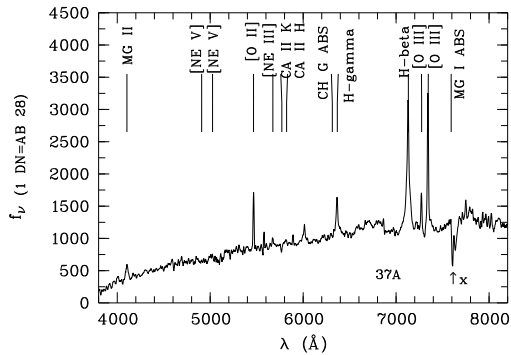,bbllx=45pt,bblly=52pt,bburx=390pt,bbury=566pt,width=6.7cm,angle=-90}
\end{minipage}
\end{center}
\caption{Keck LRIS spectrum of the NLS1 galaxy 37A ($z=0.467$). The symbol "x" marks a 
region of an atmospheric absorption band ($\sim$7600 \AA ).}
\end{figure}

The fraction of NLS1 in X-ray selected AGN samples (see for an overview Grupe in these proceedings) 
has to be considered with caution. NLS1 at redshifts above 0.65 could be
missed due to the limited wavelength coverage of the optical spectra ($\lambda
< $ 8200 \AA ). The H$\beta$/[O~III] region is covered in 13 of 69
spectroscopically identified AGNs from the ROSAT Deep Surveys. Most of our
AGNs show only broad UV emission lines, eg., C~III] $\lambda$1908 or Mg II
$\lambda$2798, at high redshifts. Recently, Rodriguez-Pascual et al. (1997)
have detected broad components of strong UV emission lines (eg., L$\alpha$
$\lambda$1216 and C~IV~$\lambda$1548) in several NLS1. If this is a common
property we can not exclude a significantly larger fraction of NLS1 in
high-redshift AGN samples.

But even the fraction of NLS1 in some low-redshift AGN samples, eg. 7\% of the ROSAT Bright 
Survey (Schwope et al. 2000) or 22\% of the RIXOS AGN sample (Puchnarewicz et
al. 1997), could result from a different classification scheme of NLS1 objects
(upper limit for FWHM$_{H\beta} < 1500$ or 2000 km s$^{-1}$). Without 
well-controlled samples at low and high redshifts it is not possible to obtain
reliable statements about the cosmological evolution of NLS1.

The fraction of NLS1 found in the ROSAT Deep Surveys ($\sim$1\%) may indicate a marginal 
contribution of NLS1 to the soft X-ray background. Nevertheless we have to
consider this as a lower limit so far.

% -------------------------------------------------------------------------

% -------------------------------------------------------------------------

\end{document}